\documentclass[journal]{IEEEtran}
\usepackage{mathtools}
\usepackage{amsmath}
\usepackage{breqn}
\usepackage{tabularx}
\usepackage{tabu}
\usepackage{booktabs}
\usepackage{tikz}
\usepackage{subfig}
\usetikzlibrary{shapes,arrows,positioning,calc}
\usepackage[noadjust]{cite}
\ifCLASSINFOpdf
\else
\fi
\usepackage{fixltx2e}
\begin{document}

\tikzset{
block/.style = {draw, fill=white, rectangle, minimum height=2.5em, minimum width=2.5em},
block2/.style = {draw, fill=white, rectangle, minimum height=2cm, minimum width=1cm},
lblock/.style = {draw, fill=white, rectangle, minimum height=8em, minimum width=3em},
tmp/.style  = {coordinate}, 
point/.style  = {coordinate},
sum/.style= {draw, fill=white, circle, node distance=1cm},
input/.style = {coordinate},
output/.style= {coordinate},
pinstyle/.style = {pin edge={to-,thin,black}
}
}
\title{Variable Speed Hydropower Plant\\with Virtual Inertia Control\\for Provision of Fast Frequency Reserves}
%
%
%

\author{Tor~Inge~Reigstad,~
        Kjetil~Uhlen
\thanks{T.I. Reigstad and K. Uhlen are with Department for Electric Power Engineering, Norwegian University of Science and Technology (NTNU), NO-7491 Trondheim, Norway (email: tor.inge.reigstad@ntnu.no, kjetil.uhlen@ntnu.no)}
}

\maketitle

\begin{abstract}

In this paper, five virtual inertia control structures are implemented and tested on a variable speed hydropower (VSHP) plant. The results show that all five can deliver fast power reserves to maintain grid stability after disturbances after a disturbance. The VSHP is well suited for the purposed since its output power can be changed almost instantaneously by utilizing the rotational energy of the turbine and generator. This will cause the turbine rotational speed to deviate from its optimal value temporarily. Then the governor control will regain the turbine rotational speed by controlling the guide vane opening and thereby the turbine flow and mechanical power. With that, the VSHP output power can be changed permanently to contribute with primarily frequency reserves.

Dynamic and eigenvalue analyses are performed to compare five different versions of the basic VSG and VSM control structures; VSG, power-frequency PID-controller with permanent droop (VSG-PID), VSM, VSM with power-frequency PD-controller (VSM-PD), and VSM with power-frequency PID-controller and permanent droop (VSM-PID). They are evaluated by two main criteria; their ability to deliver instantaneous power (inertia) to reduce the rate of change of frequency (ROCOF) and their contribution to frequency containment control (steady-state frequency droop response). 
 
\end{abstract}


%
\IEEEpeerreviewmaketitle

\section{Introduction}
%
%
%
%

\IEEEPARstart{T}{he} increasing share of non-controllable renewable energy enforces the introduction of new flexible producers and consumers to ensure the balance of the grid. In the Nordic grid, large hydro and thermal power plants have up til now supplied inertia and frequency control. The introduction of wind, solar and HVDC connection to Europa creates new production scenarios where the inertia in the grid is very low. This will compromise the frequency stability as the inverter-based generation does not provide any response to frequency deviations unless virtual inertia control is implemented. The goal of the inertia control is to control the converters to increase the inertia in the grid. Different typologies are reviewed in \cite{tamrakar2017virtual} and they are divided into three categories. The simplest type is the frequency-response based models where the power is controlled proportionally to the frequency deviation and/or the derivative of the frequency \cite{ullah2008temporary}. Other models emulate a synchronous machine by a machine model and are therefore referred to synchronous generator based models. They might contain models of inertia, damping and voltage \cite{torres2014self}. The swing-equation models are similar to the synchronous generator based models, however, they are based on a simpler power-frequency swing equation \cite{sakimoto2011stabilization}.

In this paper, Variable Speed Hydropower (VSHP) is investigated to provide Fast Frequency Reserves (FFR) and Virtual Inertia (VI) to the grid. These ancillary services can increase the frequency stability by responding to frequency deviations within $1s$ and thereby contribute to balancing the grid, maintaining the power system security and thereby improving the grid stability. As the system inertia is reduced as the share of renewable energy increases, the utilization of VI may be essential for the green shift.

The other main advantages of variable speed hydropower are increase efficiency at low production and the possibility to control the power in pumping modes while keeping the efficiency at an acceptable level \cite{valavi2016variable}. Besides, the converter technology of the VSHP can improve the speed of the voltage control and potentially increase the reactive power capability. The drawbacks are the power losses of the converters, increased costs and the reduced reliability if bypassing the converters is impossible. The limited short circuit current of the converters may also cause challenges for the generator and grid protection.

VSHP plants are particularly suited for VI control since the rotational energy of the turbine and generator can be utilized by allowing the rotational speed to deviate from its best operating point. Then the guide vane opening, water flow and thereby the mechanical power can be controlled to regain the rotational speed of the turbine. Despite other sources of virtual inertia have not the same properties, implementation of VI in VSHP is not investigated in the literature. Photovoltaic systems (PV) have very limited energy storage while the rotational speed of wind turbines \cite{wang2015inertial, diaz2014participation} must be regained by reducing the power output, which will cause additional frequency drop. Wind turbines are therefore best fitted for frequency-power response-based control with temporary grid support \cite{saarinen2018linear}. In addition, virtual inertia from batteries \cite{soni2013improvement}, capacitors \cite{arani2013implementing} and HVDC \cite{zhu2013inertia} are investigated in literature.

The power reserves are divided into four levels in \cite{diaz2014participation,entso2009p1}, each of them necessary for maintaining the balance between power generation and power consumption and thereby ensuring that the frequency is kept within the limits given by the grid codes. The fastest power reserves are the instantaneous power reserves, also called inertia. They generated by the physical stabilizing effect of all the grid-connected synchronous machines due to the energy in the rotating masses in turbines and generators and are most important the first few second after a disturbance. The frequency containment reserves (FRC) are automatically and fully activated within $30s$ in the Nordic grid. The FRC is locally activated and implemented in the governor control as frequency droop control. There are two levels of FCR in the Nordic grid, FCR-N is activated at frequency deviations $\pm 0.1 Hz$ while FCR-D is activated at $49.9 Hz$ and fully activated at $49.5 Hz$.

The two slowest levels of power reserves, Secondary and tertiary reserves, are not considered in this paper. 

A report from the Nordic TSOs \cite{orum2015future,orum2018future} states that the FFR is both the best technical and economical solution to increase the frequency stability. FRR is compared to other actions, for instance increasing the inertia by VI control \cite{statnett2018fast}. The FFR should be activated at $49.60 Hz$, have to reach the full value within 2 sec and hold this value for at least $30s$. The system should be able to react to a new frequency deviation after $15min$. An FFR marked is tested in a pilot project, however, the Pelton and Francis turbines were found to slow to deliver the power step within $2s$.

This paper investigates the instantaneous and primary power reserves of a VSHP and aims to find the best-suited control scheme for VSHP considering the defined control objectives. The paper built on paper3, however, new models and results are included, including the eigenvalue analysis.

The paper is outlined as follows: The virtual inertia models and the VSHP and grid models are presented in respectively Section \ref{VirtualInertiaModels} and \ref{Models}. The dynamic analysis results and discussion are given in Section \ref{ch:dynamic}. Section \ref{ch:eig} presents the results and discussion from the eigenvalue analysis. The conclusions are summed up in Section \ref{Conclusion}.

 

\section{Virtual Inertia Models}\label{VirtualInertiaModels}
 
\subsection{Control Objectives}

The control objectives for a VSHP are presented in \cite{2020arXiv200306298R} and are divided into objectives for internal control and grid support:
\begin{itemize}
    \item Objectives for internal control of the plant:
    \begin{itemize}
        \item To optimize the rotational speed of the turbine with respect to the efficiency,
        \item to minimize water hammering and mass oscillations,
        \item to minimize guide vane servo operation,
        \item and to minimize the hydraulic and electric losses
    \end{itemize}
    \item Objectives for grid support control:
    \begin{itemize}
        \item Contribute to FCR by faster and more precise frequency droop control,
        \item contribute to increasing the effective system inertia by virtual inertia control,
        \item improve the voltage control response time,
        \item and increase the damping in the system.
    \end{itemize}
\end{itemize}

The main focus of this paper is to maximize the grid support from the VSHP by utilizing the turbine and generator rotational energy. The VI-controllers should deliver both virtual inertia by changing the power instantaneously to reduce the rate of change of frequency (ROCOF). Besides, the VI-controller will contribute with primary reserves/FCR to regain the grid frequency as fast as possible after a disturbance.

This section presents five different virtual inertia typologies. The power-frequency PD controller known as virtual synchronous generator (VSG) \cite{van2009grid} and the virtual synchronous machine (VSM) \cite{hesse2009micro, mo2017evaluation} are known from literature. The other three typologies are extended versions of these presented in this paper; Power-frequency PID controller with permanent droop (VSG-PID), VSM with power-frequency PD controller (VSM-PD), and VSM with power-frequency PID controller and permanent droop (VSM-PID). The parameters are given in Table \ref{table_parameters} in Appendix \ref{app1}.

\subsection{Virtual Synchronous Generator}

The VSG is a power-frequency response based virtual inertia system. It tries to emulate the inertial response characteristics of a synchronous generator simply, without incorporating all the detailed equations involved in an SG. A PD controller calculates the current reference in the d-axis, $i_{g,d}^{*}$ from the deviation in grid frequency $\Delta \omega_{g}$ as shown in \eqref{eqvsg2}. The power reference $p_{g}^{*}$ is added to achieve the wanted power at zero frequency deviation and the controller compensates for deviations in voltage and for reactive power delivery \cite{tamrakar2017virtual,wang2013high}.

\begin{equation}\label{eqvsg2}
\begin{split}
    p_{vsg} &= k_{vsg,p} \Delta \omega_{g} +  \frac{k_{vsg,d} \omega_{vsg} s}{s+\omega_{vsg}} \Delta \omega_{g}  + p_{g}^{*}\\
    \Delta \omega_{g} &= \omega_g^* - \omega_g\\
    i_{g,d}^{*} &=  \frac{v_{c,d} p_{vsg} - v_{c,q} q_g}{v_{c,d}^2+v_{c,q}^2}    
\end{split}
\end{equation}

The VSG is current-controlled and not able to operate in an islanded system. Over-current protection is easily implemented, however, multiple units as current sources and the use of PLL may result in instability.

\subsection{Power-Frequency PID-controller with Permanent Droop}

An alternative control layout for the VSG with PID controller and permanent droop (VSG-PID) is proposed in paper3, as shown in \eqref{eqvsgpid}. The benefit of the PID controller is that it can be tuned to be some faster than the PD-controller. Due to the integration part, the permanent droop is needed to ensure power power-sharing between generators as with conventional power plants.


\begin{equation}\label{eqvsgpid}
\begin{split}
    p_{vsg-pid} &= k_{vsg-pid,p} \epsilon + \frac{k_{vsg-pid,d} \omega_{vsg-pid} s}{s+\omega_{vsg-pid}} \epsilon \\
    & \quad + \frac{k_{vsg-pid,i}}{s}  \epsilon + p_{g}^{*}\\
    \epsilon &= \omega_g^* - \omega_g - R p_f\\
    i_{g,d}^{*} &=  \frac{v_{c,d} p_{vsg-pid} - v_{c,q} q_g}{v_{c,d}^2+v_{c,q}^2}\\
\end{split}
\end{equation}


\subsection{Virtual Synchronous Machines}

\begin{figure*}[!t]
\centering
\begin{tikzpicture}[auto, node distance=1cm,>=latex']

    \node [sum] (sum1){};
    \node [input, left of=sum1, node distance =4.2cm] (inputvref) {};
    \draw [->] (inputvref) -- node[anchor=south,pos=0.05]{$v_g^*$} (sum1);
    \node [point, above of =sum1,node distance=1cm] (sum1a){};
    \node [block, left of=sum1a,node distance=2.0cm] (abs) {$\mid x \mid$};
    \draw [->] (abs) -| node[pos=0.9] {$-$} (sum1);
    \node [input, left of=abs, node distance =2.2cm] (inputv0) {};
    \draw [->] (inputv0) -- node[anchor=south,pos=0.1]{$\mathbf{v}_{g}$} (abs);
    \node [point, below of =sum1,node distance=1cm] (sum1b){};
    \node [block, left of=sum1b,node distance=1.0cm] (kq) {$k_q$};
    \node [sum, left of =kq,node distance=1cm] (sum2){};
    \draw [->] (sum2) --  (kq);
    \draw [->] (kq) -|  (sum1);
    \node [point, below of =sum2,node distance=1cm] (sum2b){};
    \node [block, left of=sum2b,node distance=1.0cm] (wqf) {$\frac{w_{qf}}{s+w_{qf}}$};
    \node [input, left of=sum2, node distance =2.2cm] (inputqref) {};
    \draw [->] (inputqref) -- node[anchor=south,pos=0.1]{$q_g^*$} (sum2);
    \node [input, left of=wqf, node distance =1.2cm] (inputq) {};
    \draw [->] (inputq) -- node[anchor=south,pos=0.35]{$q_g$} (wqf);
    \draw [->] (wqf) -| node[pos=0.9] {$-$} (sum2);
    
    \node [block, right of=sum1, node distance=1.0cm] (PI) {PI};
    \draw [->] (sum1) -- (PI);
    \node [sum, right of =PI,node distance=1cm] (sum3){};
    \draw [->] (PI) --  (sum3);
    \node [block, right of=sum1a, node distance=1.0cm] (k) {$k_{ffe}$};
    \draw [->] (sum1) -- (PI);
    \draw [->] (sum1a) -- (k);
    \draw [->] (k) -| (sum3);
    
    \node [sum, right of =sum3,node distance=3.6cm] (sum4){};
    \draw [->] (sum3) -- node[anchor=south,pos=0.5]{$\hat{v}_e$} (sum4);
    \node [point, below of =sum4,node distance=1cm] (sum4b){};
    
    \node [block, left of=sum4b,node distance=1.0cm] (wvf) {$\frac{w_{vf}}{s+w_{vf}}$};
    \node [input, left of=wvf, node distance =1.5cm] (inputv0) {};
    \draw [->] (inputv0) -- node[anchor=south,pos=0.1]{$\mathbf{v}_0$} (wvf);
    \draw [->] (wvf) -| node[pos=0.9] {$-$}  (sum4);
    
    \node [point, right of =sum4,node distance=1cm] (sum4r){};
    \node [block,below of =sum4r, node distance =0.2cm] (pd){$\times/\div$};
    \draw [->] (sum4) --   ([yshift=0.2cm]pd.west);
    
    \node [block, below of=wvf,node distance=1.0cm] (rs) {$r_s + j x l_s$};
    \node [input, left of=rs, node distance =1.5cm] (inputw) {};
    \node [point, right of =rs,node distance=1.2cm] (rsr){};
    \draw [->] (rs) -- (rsr) |-  ([yshift=-0.2cm]pd.west);
    
    \node [output, right of=pd, node distance=1.6cm] (outis) {};
    \draw [->] (pd) -- node[anchor=south,pos=0.7]{$\mathbf{i}_s=\mathbf{i}_{cv}$}(outis);
    
    \node [sum, below of =wqf,node distance=2cm] (sum5){};
    \node [input, left of=sum5, node distance =1.2cm] (inputwref) {};
    \node [input, below of=inputwref, node distance =1.0cm] (inputw2) {};
    \draw [->] (inputwref) -- node[anchor=south,pos=0.2]{$\omega_g^*$} (sum5);
    \draw [->] (inputw2) -| node[anchor=south,pos=0.1]{$\omega_{vsm}$} node[pos=0.9] {$-$}  (sum5);
    
    \node [block, right of=sum5, node distance=1.0cm] (kw) {$k_w$};
    \draw [->] (sum5) -- (kw);
    \node [sum, right of =kw,node distance=1cm] (sum6){};
    \draw [->] (kw) --  (sum6);
    \node [input, below of=inputw2, node distance =1.0cm] (inputpref) {};
    \draw [->] (inputpref) -| node[anchor=south,pos=0.1]{$p_{g,r}^* = p_g^*$} (sum6);
    \node [sum, right of =sum6,node distance=2cm] (sum7){};
    \draw [->] (sum6) -- node[anchor=south,pos=0.5]{$p_r^*$} (sum7);
    \node [point, above of =sum7,node distance=0.8cm] (sum7a){};
    \node [input, left of=sum7a, node distance=1.0cm] (inputp) {};
    \draw [->] (inputp) -| node[anchor=south,pos=0.05]{$p_g$} node[pos=0.9] {$-$}(sum7);
    \node [block, right of=sum7,node distance=2.0cm] (ta) {$\frac{1}{T_a s}$};
    \draw [->] (sum7) --  (ta);
    \node [sum, right of =ta,node distance=2.5cm] (sum8){};
    \draw [->] (ta) --  (sum8);
    \node [sum, below of =ta,node distance=1.0cm] (sum9){};
    \node [block, left of=sum9, node distance=1.0cm] (kd) {$k_d$};
    \draw [->] (sum9) --  (kd);
    \draw [->] (kd) -| node[pos=0.92] {$-$} (sum7);
    \node [point, left of =sum8,node distance=0.9cm] (sum8l){};
    \draw [->] (sum8l) |-  (sum9);
    \node [point, below of =sum8l,node distance=1.0cm] (sum8lb){};
    \node [point, below of =sum8l,node distance=2.0cm] (sum8lb2){};
    \node [block, left of=sum8lb2,node distance=0.8cm] (wd) {$\frac{\omega_d}{s+\omega_d }$};
    \draw [->] (sum8lb) |-  (wd);
    \draw [->] (wd) -| node[pos=0.9] {$-$} (sum9);
    \node [block, below of=sum8, node distance=1.0cm] (k) {1};
    \draw [->] (k) --  node[pos=0.8] {$-$} (sum8);
    \node [block, right of=sum8,node distance=1.5cm] (pi) {$\frac{\omega_b}{s}$};
    \draw [->] (sum8) --  (pi);
    \node [output, right of=pi, node distance=1.4cm] (outtheta) {};
    \draw [->] (pi) -- node[anchor=south,pos=0.7]{$\theta_{vsm}$} (outtheta);
    \node [output, above of=outtheta, node distance=0.8cm] (outw) {};
    \draw [->] (sum8l) |- node[anchor=south,pos=0.95]{$\omega_{vsm}$} (outw);
    \node [point, above of =sum8l,node distance=0.8cm] (sum8la){};
    \draw [->] (sum8la) -| (inputw) -- node[anchor=south,pos=0.1]{$\omega_{vsm}$} (rs);
    
    \draw[red,thick,dotted]  ($(inputq)+(-0.2,-0.6)$)  rectangle ($(sum3)+(0.5,1.6)$) ;
    \draw[blue,thick,dotted] ($(inputw)+(-0.5,-0.6)$)  rectangle ($(pd)+(2.0,1.8)$);
    \draw[green,thick,dotted] ($(inputpref)+(-0.2,-0.6)$)  rectangle ($(sum6)+(0.5,1.3)$);
    \draw[orange,thick,dotted] ($(inputp)+(-0.4,-3.4)$)  rectangle ($(outw)+(0.2,0.5)$);

    \node [align=left] at ($(PI)+(0.2,-2.3)$) {Voltage control};
    \node [align=center] at ($(sum4)+(-1.5,1.3)$) {Electrical model};
    \node [align=left] at  ($(kw)+(-0.8,1.0)$) {Frequency control};
    \node [align=left] at  ($(wd)+(-2.9,-0.3)$) {Inertia model};

\end{tikzpicture}
\caption{Virtual synchronous machine (VSM)} \label{figVSM}
\end{figure*}
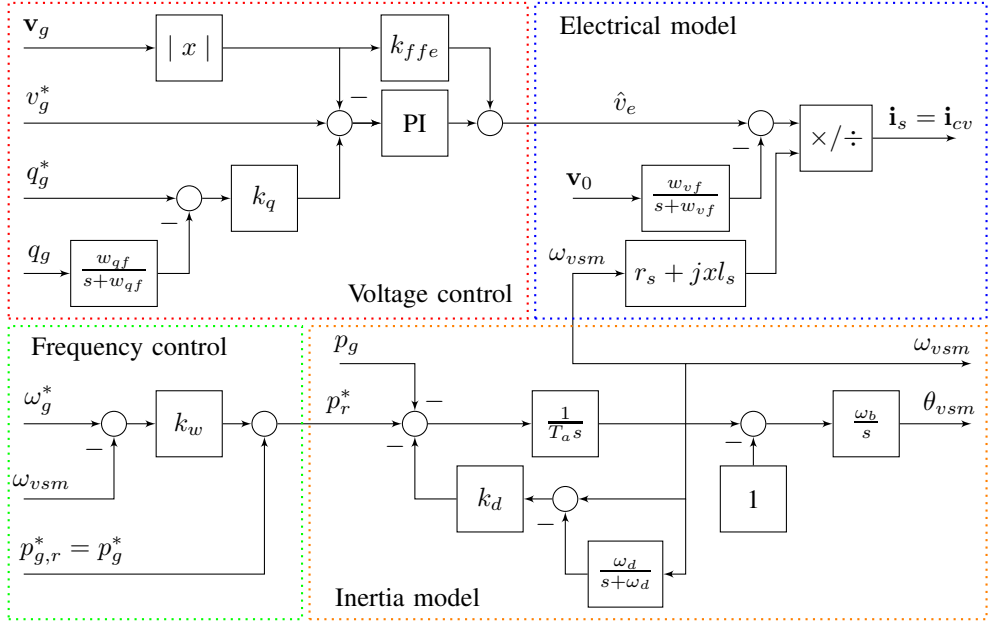

The VSM is a synchronous generator based virtual inertia model. In this paper, the model presented in \cite{mo2017evaluation} is utilized. It includes models for voltage control, frequency control and model for inertia and the electrical system as shown in Fig. \ref{figVSM}. The main benefit of the VSM is that it can work in islanded systems without changing parameters and control structure. \cite{beck2007virtual,driesenvirtual}

\subsection{VSM with Power-Frequency PD controller}

The main drawback with the VSM is that its output power does return relatively quickly to the reference power, even there are still deviations in the grid frequency. It does therefore not contribute to primary control/FRC. This problem can be solved by combining the VSM with other frequency regulation schemes. 

The first option to be tested is to add the output power reference from a power-frequency PD controller to the VSM virtual inertia $p_{g,r}^*$ as presented in \eqref{eqvsmvsg}. Both a deviation and a change in grid frequency will adjust the power reference to the VSM and the VSHP will contribute to power reference. A PLL is used to measure the grid frequency $\omega_g$.

\begin{equation}\label{eqvsmvsg}
\begin{split}
    p_r^* &= p_g^* + k_{\omega} \left( \omega_{vsm}^* -\omega_{vsm} \right) + p_{vsm-pd}\\
    p_{vsm-pd} &= k_{vsm-pd,p} \Delta \omega_{g} +  \frac{k_{vsm-pd,d} \omega_{vsm-pd} s}{s+\omega_{vsm-pd}} \Delta \omega_{g} \\
    \Delta \omega_{g} &= \omega_g^* - \omega_g\\
\end{split}
\end{equation}

\subsection{VSM with Power-Frequency PID controller and Permanent Droop}

Frequency control can alternatively be added to the VSM by including a PID-controller with permanent droop as presented in \eqref{eqvsmpid}. The output of this controller (VSM-PID) is added to the virtual power $p_{g,r}^*$. The function of the PID-controller will be similar to the function of the governor of a conventional hydropower plant. However, since the speed of the governor servo is not limiting, the frequency response of the VSM-PID will be significantly faster primary frequency control.


\begin{equation}\label{eqvsmpid}
\begin{split}
    p_r^* &= p_g^* + k_{\omega} \left( \omega_{vsm}^* -\omega_{vsm} \right) + p_{vsm-pid}\\
    p_{vsm-pid} &= k_{vsm-pid,p} \epsilon + \frac{k_{vsm-pid,d} \omega_{vsm-pid} s}{s+\omega_{vsm-pid}} \epsilon \\
    & \quad + \frac{k_{vsm-pid,i}}{s}  \epsilon \\
    \epsilon &= \omega_g^* - \omega_g - R p_f\\
\end{split}
\end{equation}

$p_f$ is the low-pass filtered active output power $p_g$.

\section{Variable Speed Hydropower and Grid Models}\label{Models}

The VI-controllers are tested on the VSHP model and two-area power system presented in \cite{reigstad2019variable} with some modifications. The grid converter outer control loop is replaced by the VI-controllers. For the VSG controllers, only the active power controller is replaced and the VSG supplies the current reference in the d-axis $i_{g,d}^*$ to the current controller. The reactive power controller is kept. For the VSM controllers, both the active and reactive power controllers are replaced by the VSM model. 

\section{Dynamic Analysis}\label{ch:dynamic}

\begin{figure}[!t]
\centering
    \includegraphics[scale=0.8]{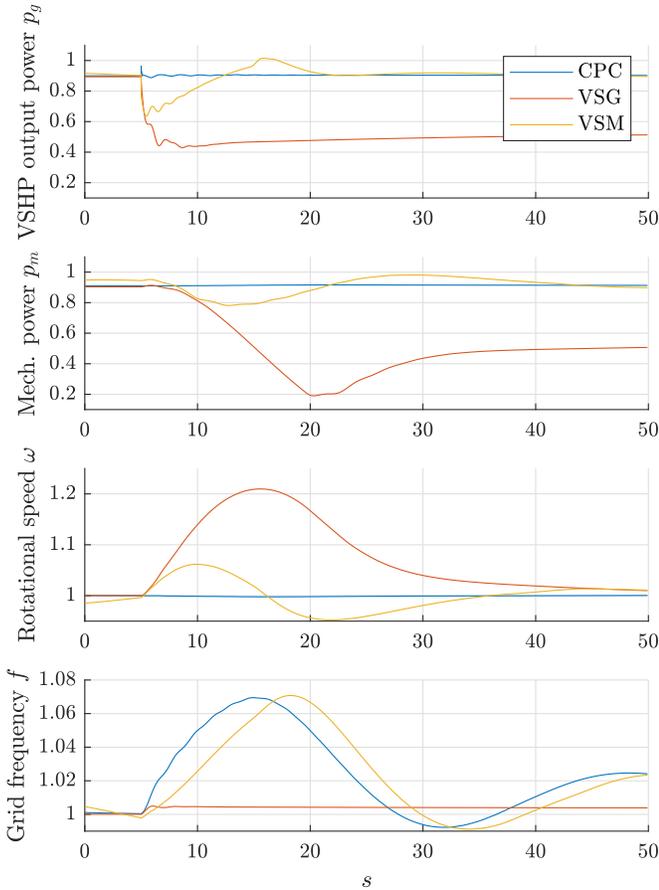}
    \caption{Comparison of load step responses on Bus 7 for different VSHP control schemes} \label{fig711}
\end{figure}

The performance of the different control schemes is first analyzed by dynamic simulations, both in cases with overproduction and underproduction in the grid. In Fig. \ref{fig711} the responses to step load loss at Bus 7 is shown. The responses illustrate the difference between a constant power controller (CPC), the VSM that provides an inertial response and the VSG, providing both inertial response and frequency containment. 

The VSG contributes with FCR since the droop characteristic of the VSG controller will cause the output power to stabilize at a lower value after the disturbance. With the VSM controller, the output power is returning to its reference value since it has zero steady-state feedback from grid frequency. The maximal frequency deviation will be similar, or even higher, than with the constant power controller. The performance of the VSM can be improved by controlling the power reference to the VSM with a PD controller with feedback from grid frequency (VSM-PD) or a PID controller with permanent frequency droop (VSM-PID). 

\begin{figure}[!t]
\centering
    \includegraphics[scale=0.8]{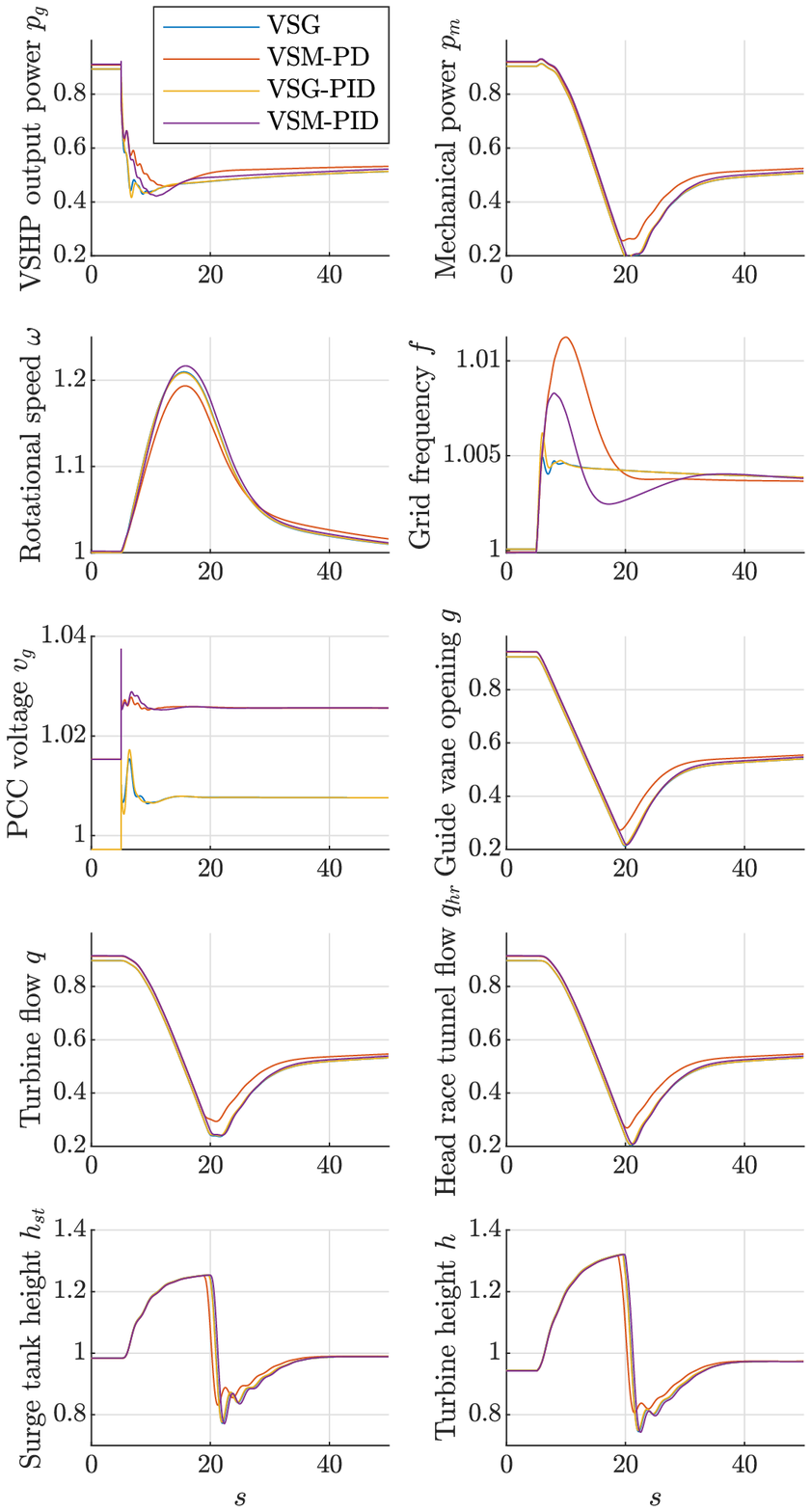}
    \caption{Comparison of load step responses on Bus 7 for different VSHP control schemes} \label{figStepBus7}
\end{figure}

Fig. \ref{figStepBus7} shows the results for both the electric and hydraulic variables of the four most promising controllers, the VSG, VSG-PID, VSM-PD and VSM-PID. The VSM is not included since it does not have droop control and does not contribute to FCR. The main observation is that the VSG and the VSG-PID have the shortest response time and can reduce the maximum deviation in grid frequency $f$. This is due to a larger power reduction of the VSHP output power $p_g$ with VSG-based inertia controllers compared to the VSM-based inertia controllers from $1-10s$ after the disturbance.

The oscillation in VSHP output power $p_g$ is some larger for the VSG-based inertia, however, they are small and well-damped. As seen in Fig. \ref{figStepBus7}, the variables of the hydraulic system are mostly not affected by the choice of VI-controllers. The VSM-PC controller stands out because of its slower reaction to grid frequency deviations. The reduction in VSHP output power $p_g$ is less, causing less deviations in turbine rotational speed $\omega$, turbine power $p_m$, guide vane opening $g$ and thereby turbine flow $q$ and surge tank head $h_{st}$.

\begin{figure}[!t]
\centering
    \includegraphics[scale=0.8]{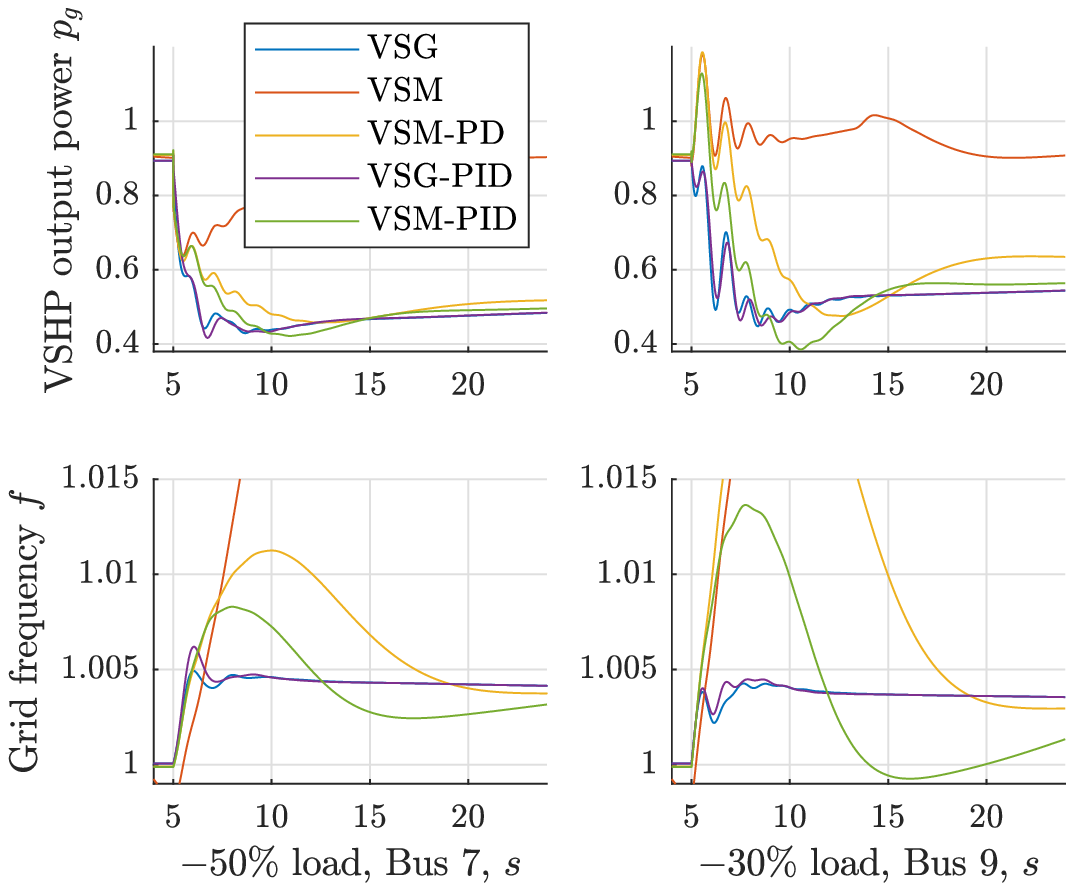}
    \caption{Step response with load reduction at respectively $50\%$ on Bus 7 and $30\%$ on Bus 9} \label{figStep}
\end{figure}

\begin{figure}[!t]
\centering
    \includegraphics[scale=0.8]{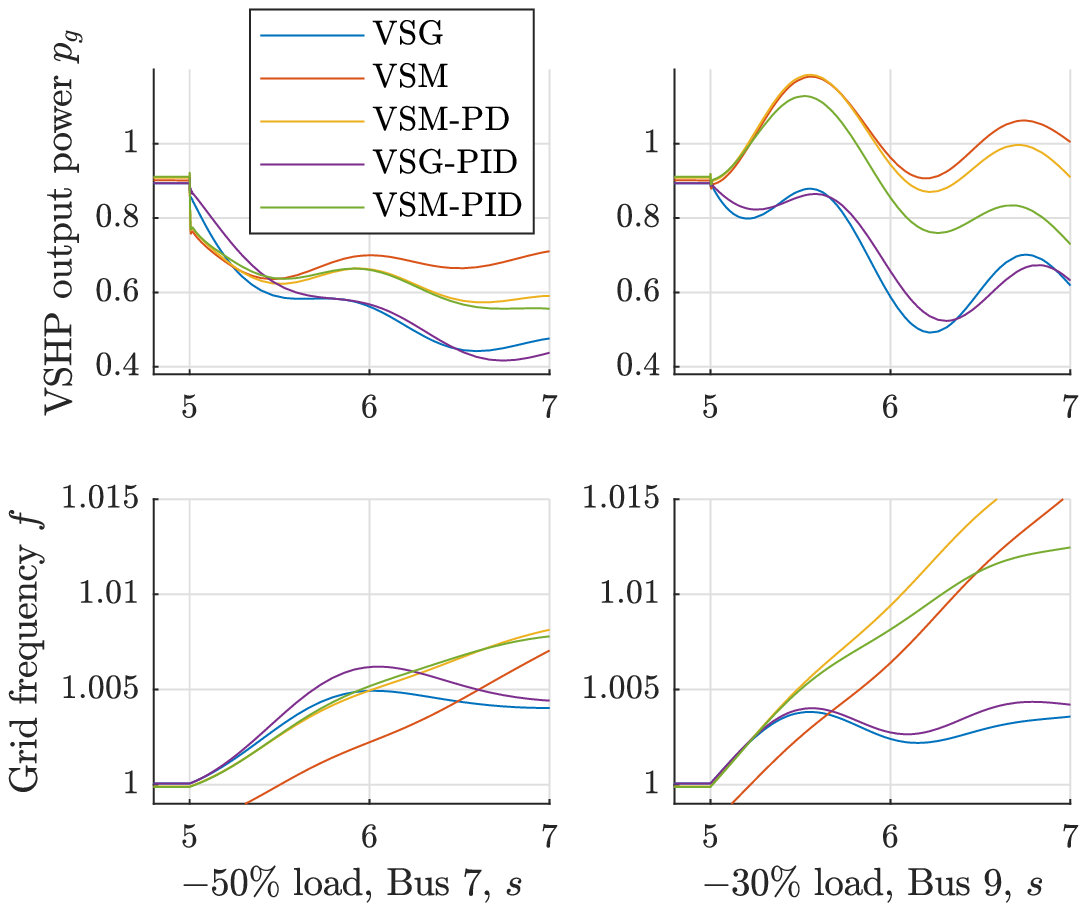}
    \caption{Step response with load reduction at respectively $50\%$ on Bus 7 and $30\%$ on Bus 9} \label{figStepZoomed}
\end{figure}

Two cases with load loss at different locations are compared in Fig. \ref{figStep}, showing the first 25 sec after the load loss and Fig. \ref{figStepZoomed} is zoomed in on the first 2 sec to show the difference in inertia response. The sizes of the load losses are similar and they are located in respective close to the VSHP (Bus 7 in Area 1) and far away from the VSHP (Bus 9 in Area 2).

From Fig. \ref{figStepZoomed}, we observe that the output power of the VSM-based VI-controllers do have the fastest respond the first milliseconds after the disturbance when the load loss is close to the VSHP. They do therefore deliver more inertia than the VSG-based VI-controllers. However, when the load loss appears at Bus 9 farther away from the VSHP, the situation is totally different, as discussed below.

Since most of the FRC is delivered by the VSHP, the power from Area 1 to Area 2 will decrease in the case of load loss in Area 2 and trigger power oscillations between the two areas. Due to these power oscillations, there will by higher oscillations in both the output power of the VSG- and VSM-based VI-controllers when the load loss is far away from the VSHP. This is related to inter-area power oscillation between the two areas of the system, as also observed in the nearby generators. Particularly, these power oscillations affect the VSM-based VI-controllers since they are dependent on the voltage angle. In the case of load loss at Bus 9, the power output of the VSM-based VI-controllers actually increases right after the load loss, causing a higher deviation in grid frequency $f$. The dependence on the voltage seems to be a major disadvantage by emulating a synchronous machine and the use of VSM might cause problems in the system with large power oscillations.  

The VSG-based VI-controllers do only consider the frequency, and not the angle when controlling the VSHP output power. Therefore, the power oscillation will have less impact and the response to a load loss in Area 2 will be almost as fast as if the load loss occurred in Area 1, however with more oscillations. Besides, the VSG-based VI-controllers reduces the frequency deviation by a larger reduction in VSHP output power between $1-3s$ after the load loss. 

\section{Eigenvalue Analysis}\label{ch:eig}

\begin{figure}[!t]
\centering
    \includegraphics[scale=0.8]{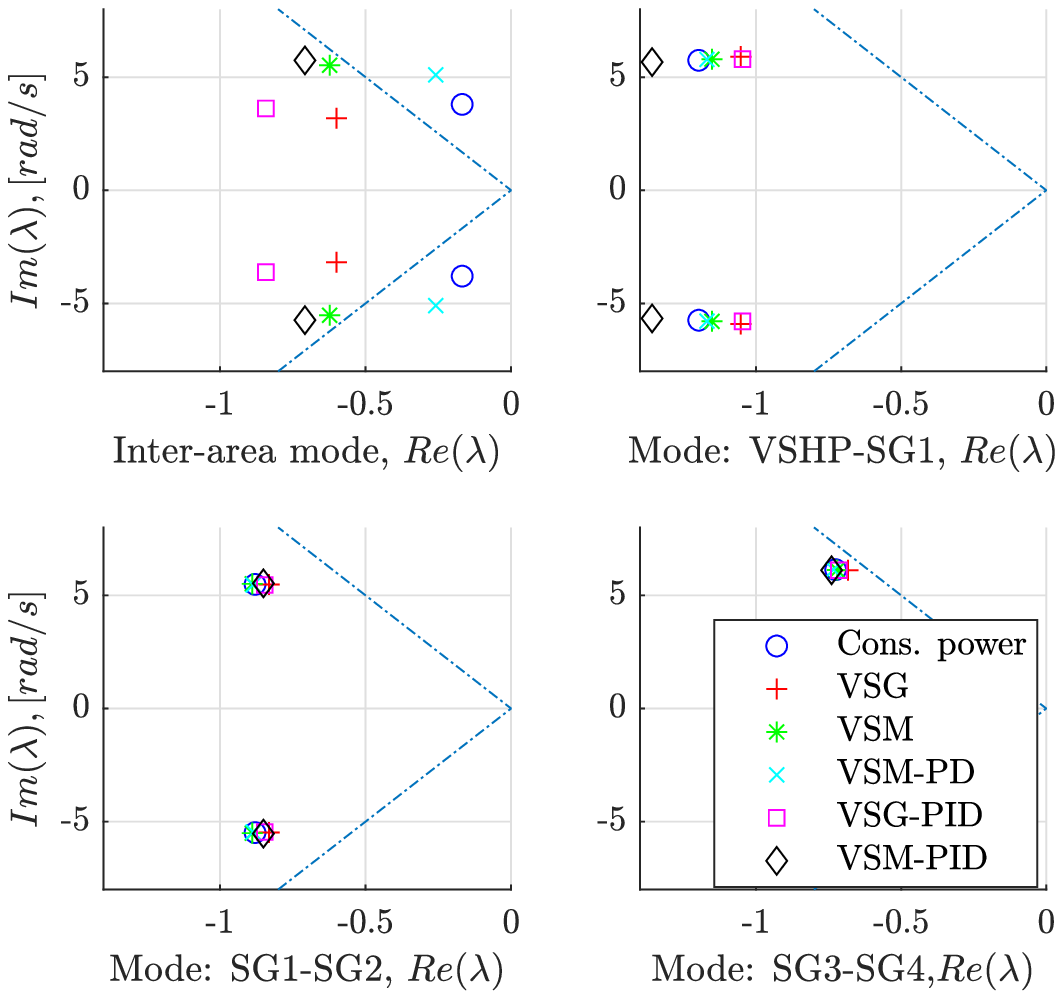}
    \caption{Power oscillation modes for different VSHP control schemes} \label{figPO}
\end{figure}

In this chapter, the most important results from the eigenvalue analysis are presented. In the Kundur Two-Area system, there exists an interarea mode between Area 1 and Area 2 and a local mode in each area, between respectively SG1-SG2 and SG3-SG4. As the VSHP is connected on Bus 5, close to SG1, a new local mode appears between the VSHP and SG1. Fig. \ref{figPO} shows these four modes for the different control schemes of the VSHP.

The inter-area mode is the mode most dependent on the VSHP control scheme. Compared to the case with constant power control, the modes of the VSM and its variants have higher frequency and higher relative damping. The damping of this mode for the VSM-PD is still poor. The relative damping of the inter-area mode is doubled when the VSG control of the VSHP is introduced. In addition, the frequency of the oscillation is reduced. The relative damping is slightly better for the VSG-PID than for the VSG. 

The relative damping of the mode between the VSHP and SG1 is slightly better for the VSM based control schemes than the VSG based control schemes. The local modes between the generators are more or less independent of the VSHP control scheme.

\section{Conclusion}\label{Conclusion}

Two VI controllers, the VSG and the VSM are developed further in this paper to increase the contribution of instantaneously and primarily frequency resources, which are the main objectives for grid support control. At the same time, the objectives of internal control of the VSHP has to be considered to limit water hammering, mass oscillations and guide vane operation and to ensure that the turbine rotational speed regains within an acceptable time.

The VSM topology shows the fastest response when simulating a disturbance and thus delivers the best inertial response. However, since it has zero steady-state feedback from grid frequency, the output power returns to its reference value and the VSM is not contributing to the frequency control. Frequency control is added by controlling the power reference to the VSM with a PD controller with feedback from grid frequency (VSM-PD) or a PID controller with permanent frequency droop (VSM-PID).

The VSG-based controller typologies do not have an instantaneous response to the load loss, however, the power response is larger from 200 ms to 5 seconds after the disturbance. In addition, the VSG-based controllers damps oscillations against other generators better, resulting in lower frequency deviation. 

Although the VSM controller shows the fastest response during the first 200 ms after a disturbance, the VSG controller provides better frequency regulation for the next 5 seconds. In cases where the disturbance is far away from the VSHP, the performance of the VSM controller is reduced and the VSG perform better regarding the instantaneous response (synthetic inertia) and permanent frequency droop control. 

The dynamic analysis clearly shows that the VSG has the best performance from a grid-integration point of view. However, since the VSM controller is based on emulating the response of a synchronous generator, the PLL is not needed. This makes the VSM able to work in islanded systems without changing parameters and control structure.

The transient behaviour of the hydraulic system is more or less equal for the VSG, VSG-PID and VSM-PID. The VSHP output power response of the VSM-PD is marginal smaller, causing smaller deviations in the hydraulic system variables.

\appendices
\section{Parameters, set-points and variable}\label{app1}

The VI model parameters are given in Table \ref{table_parameters}.

\begin{table}[!t]
\renewcommand{\arraystretch}{1.3}
\caption{Parameters and set-points}
\label{table_parameters}
\centering
\begin{tabularx}{0.85\linewidth}{|Xr|}
\hline
\textbf{Parameter} & \textbf{Value}  \\
\hline
\multicolumn{2}{|l|}{\textbf{CPC}}\\
Proportional gain $k_{Pp}$  &  0.045 p.u.\\
Integral gain $k_{Pi}$  & 0.023 p.u.\\
\multicolumn{2}{|l|}{\textbf{VSG}}\\
Proportional gain $k_{vsg,p}$  &  100 p.u.\\
Derivative gain $k_{vsg,d}$  & 33.6 p.u.\\
Derivative filter constant $\omega_{vsg}$  & 0.01 s \\
\multicolumn{2}{|l|}{\textbf{VSG-PID}}\\
Proportional gain $k_{vsg-pid,p}$  &  100 p.u.\\
Integral gain $k_{vsg-pid,i}$  &  286 p.u.\\
Derivative gain $k_{vsg-pid,d}$  & 33.6 p.u.\\
Derivative filter constant $\omega_{vsg-pd}$  & 0.01 s \\
\multicolumn{2}{|l|}{\textbf{VSM}}\\
Voltage controller proportional gain & 0.29 p.u. \\
Voltage controller integral gain & 92 p.u. \\ 
Voltage feedforward in voltage controller $k_{ffe}$ & 0 p.u.\\
Reactive power filter $\omega_{qf}$ & 200 rad/s  \\
Reactive power droop gain $k_q$ & 0.1 p.u. \\
Voltage low pass filter $\omega_{vf}$ & 200 rad/s  \\
SM inductance $l_s$ & 0.25 p.u. \\
SM resistance $r_s$ & 0.01 p.u. \\
Frequency controller gain $k_{\omega}$ & 20 p.u.\\
Inertia constant $T_a$ & 4 s \\
Damping coefficient $k_d$ & 40  p.u. \\
Damping filter $\omega_d$ &  5 rad/s  \\
Rated angular frequency $\omega_b$ & 50 rad/s \\
Active damping gain $k_{AD}$ & 0.3 p.u. \\
Active damping $\omega_{AD}$ & 50 rad/s \\
\multicolumn{2}{|l|}{\textbf{VSM-PD}}\\
Proportional gain $k_{vsm-pd,p}$  &  100 p.u.\\
Derivative gain $k_{vsm-pd,d}$  & 500 p.u.\\
Derivative filter constant $\omega_{vsm-pd}$  & 1 s \\
Frequency controller gain $k_{vsm-pd}$  & 200 p.u.\\
\multicolumn{2}{|l|}{\textbf{VSM-PID}}\\
Proportional gain $k_{vsm-pid,p}$  &  3000 p.u.\\
Integral gain $k_{vsm-pid,i}$  &  476 p.u.\\
Derivative gain $k_{vsm-pid,d}$  & 12600 p.u.\\
Derivative filter constant $\omega_{vsm-pid}$ & 1 s \\
Frequency controller gain $k_{vsm-pid}$  & 2000 p.u.\\
\multicolumn{2}{|l|}{\textbf{Common parameters}}\\
Droop $R_{d}$ & 0.01  p.u.\\
PLL frequency filter constant & 0.001 s \\
\hline
\end{tabularx}
\end{table}


\section*{Acknowledgment}

This work was supported by Norwegian Research Centre for Hydropower Technology (HydroCen).

\ifCLASSOPTIONcaptionsoff
  \newpage
\fi



\bibliographystyle{IEEEtran}
\bibliography{Paper3}

\begin{thebibliography}{10}
\providecommand{\url}[1]{#1}
\csname url@samestyle\endcsname
\providecommand{\newblock}{\relax}
\providecommand{\bibinfo}[2]{#2}
\providecommand{\BIBentrySTDinterwordspacing}{\spaceskip=0pt\relax}
\providecommand{\BIBentryALTinterwordstretchfactor}{4}
\providecommand{\BIBentryALTinterwordspacing}{\spaceskip=\fontdimen2\font plus
\BIBentryALTinterwordstretchfactor\fontdimen3\font minus
  \fontdimen4\font\relax}
\providecommand{\BIBforeignlanguage}[2]{{%
\expandafter\ifx\csname l@#1\endcsname\relax
\typeout{** WARNING: IEEEtran.bst: No hyphenation pattern has been}%
\typeout{** loaded for the language `#1'. Using the pattern for}%
\typeout{** the default language instead.}%
\else
\language=\csname l@#1\endcsname
\fi
#2}}
\providecommand{\BIBdecl}{\relax}
\BIBdecl

\bibitem{tamrakar2017virtual}
U.~Tamrakar, D.~Shrestha, M.~Maharjan, B.~P. Bhattarai, T.~M. Hansen, and
  R.~Tonkoski, ``Virtual inertia: Current trends and future directions,''
  \emph{Applied Sciences}, vol.~7, no.~7, p. 654, 2017.

\bibitem{ullah2008temporary}
N.~R. Ullah, T.~Thiringer, and D.~Karlsson, ``Temporary primary frequency
  control support by variable speed wind turbines—potential and
  applications,'' \emph{IEEE Transactions on Power Systems}, vol.~23, no.~2,
  pp. 601--612, 2008.

\bibitem{torres2014self}
M.~A. Torres~L, L.~A. Lopes, L.~A. Moran~T, and J.~R. Espinoza~C, ``Self-tuning
  virtual synchronous machine: a control strategy for energy storage systems to
  support dynamic frequency control,'' \emph{IEEE Transactions on Energy
  Conversion}, vol.~29, pp. 833--840, 2014.

\bibitem{sakimoto2011stabilization}
K.~Sakimoto, Y.~Miura, and T.~Ise, ``Stabilization of a power system with a
  distributed generator by a virtual synchronous generator function,'' in
  \emph{Power Electronics and ECCE Asia (ICPE \& ECCE), 2011 IEEE 8th
  International Conference on}.\hskip 1em plus 0.5em minus 0.4em\relax IEEE,
  2011, pp. 1498--1505.

\bibitem{valavi2016variable}
M.~Valavi and A.~Nysveen, ``Variable-speed operation of hydropower plants:
  Past, present, and future,'' in \emph{Electrical Machines (ICEM), 2016 XXII
  International Conference on}.\hskip 1em plus 0.5em minus 0.4em\relax IEEE,
  2016, pp. 640--646.

\bibitem{wang2015inertial}
S.~Wang, J.~Hu, X.~Yuan, L.~Sun \emph{et~al.}, ``On inertial dynamics of
  virtual-synchronous-controlled dfig-based wind turbines,'' \emph{IEEE Trans.
  Energy Convers}, vol.~30, no.~4, pp. 1691--1702, 2015.

\bibitem{diaz2014participation}
F.~D{\'\i}az-Gonz{\'a}lez, M.~Hau, A.~Sumper, and O.~Gomis-Bellmunt,
  ``Participation of wind power plants in system frequency control: Review of
  grid code requirements and control methods,'' \emph{Renewable and Sustainable
  Energy Reviews}, vol.~34, pp. 551--564, 2014.

\bibitem{saarinen2018linear}
L.~Saarinen, P.~Norrlund, W.~Yang, and U.~Lundin, ``Linear synthetic inertia
  for improved frequency quality and reduced hydropower wear and tear,''
  \emph{International Journal of Electrical Power \& Energy Systems}, vol.~98,
  pp. 488--495, 2018.

\bibitem{soni2013improvement}
N.~Soni, S.~Doolla, and M.~C. Chandorkar, ``Improvement of transient response
  in microgrids using virtual inertia,'' \emph{IEEE transactions on power
  delivery}, vol.~28, no.~3, pp. 1830--1838, 2013.

\bibitem{arani2013implementing}
M.~F.~M. Arani and E.~F. El-Saadany, ``Implementing virtual inertia in
  dfig-based wind power generation,'' \emph{IEEE Transactions on Power
  Systems}, vol.~28, no.~2, pp. 1373--1384, 2013.

\bibitem{zhu2013inertia}
J.~Zhu, C.~D. Booth, G.~P. Adam, A.~J. Roscoe, and C.~G. Bright, ``Inertia
  emulation control strategy for vsc-hvdc transmission systems,'' \emph{IEEE
  Trans. Power Syst}, vol.~28, no.~2, pp. 1277--1287, 2013.

\bibitem{entso2009p1}
C.~E. O.~H. ENTSO-E, ``P1-policy 1: Load-frequency control and performance,''
  2009.

\bibitem{orum2015future}
E.~{\O}rum, M.~Kuivaniemi, M.~Laasonen, A.~I. Bruseth, E.~A. Jansson,
  A.~Danell, K.~Elkington, and N.~Modig, ``Future system inertia,''
  \emph{ENTSOE, Brussels, Tech. Rep}, 2015.

\bibitem{orum2018future}
E.~{\O}rum, L.~Haarla, M.~Kuivaniemi, M.~Laasonen, A.~Jerk{\o}, I.~Stenkl{\o}v,
  F.~Wik, K.~Elkington, R.~Eriksson, N.~Modig, and P.~Schavemaker, ``Future
  system inertia 2,'' \emph{ENTSOE, Brussels, Tech. Rep}, 2018.

\bibitem{statnett2018fast}
Statnett, ``Fast frequency reserves 2018 - pilot for raske frekvensreserver,''
  \url{https://www.statnett.no/contentassets/.../evaluering-av-raske-frekvensreserver.pdf},
  Tech. Rep., 2018, accessed: 2019-02-15.

\bibitem{2020arXiv200306298R}
T.~I. {Reigstad} and K.~{Uhlen}, ``{Modelling of Variable Speed Hydropower for
  Grid Integration Studies},'' \emph{arXiv e-prints}, p. arXiv:2003.06298, Mar.
  2020.

\bibitem{van2009grid}
M.~Van~Wesenbeeck, S.~De~Haan, P.~Varela, and K.~Visscher, ``Grid tied
  converter with virtual kinetic storage,'' in \emph{PowerTech, 2009 IEEE
  Bucharest}.\hskip 1em plus 0.5em minus 0.4em\relax IEEE, 2009, pp. 1--7.

\bibitem{hesse2009micro}
R.~Hesse, D.~Turschner, and H.-P. Beck, ``Micro grid stabilization using the
  virtual synchronous machine (visma),'' in \emph{Proceedings of the
  International Conference on Renewable Energies and Power Quality
  (ICREPQ’09), Valencia, Spain}, 2009, pp. 15--17.

\bibitem{mo2017evaluation}
O.~Mo, S.~D'Arco, and J.~A. Suul, ``Evaluation of virtual synchronous machines
  with dynamic or quasi-stationary machine models,'' \emph{IEEE Transactions on
  Industrial Electronics}, vol.~64, no.~7, pp. 5952--5962, 2017.

\bibitem{wang2013high}
Y.~Wang, G.~Delille, H.~Bayem, X.~Guillaud, and B.~Francois, ``High wind power
  penetration in isolated power systems—assessment of wind inertial and
  primary frequency responses,'' \emph{IEEE Transactions on Power Systems},
  vol.~28, no.~3, pp. 2412--2420, 2013.

\bibitem{beck2007virtual}
H.-P. Beck and R.~Hesse, ``Virtual synchronous machine,'' in \emph{Electrical
  Power Quality and Utilisation, 2007. EPQU 2007. 9th International Conference
  on}.\hskip 1em plus 0.5em minus 0.4em\relax IEEE, 2007, pp. 1--6.

\bibitem{driesenvirtual}
J.~Driesen and K.~Visscher, ``Virtual synchronous generators, 2008,'' in
  \emph{Proceedings of the IEEE PES Meeting}, pp. 20--24.

\bibitem{reigstad2019variable}
T.~I. Reigstad and K.~Uhlen, ``Variable speed hydropower conversion and
  control,'' \emph{IEEE Transactions on Energy Conversion}, vol.~35, no.~1, pp.
  386--393, March 2020.

\end{thebibliography}



%

\begin{IEEEbiography}[{\includegraphics[width=1in,height=1.25in,clip,keepaspectratio]{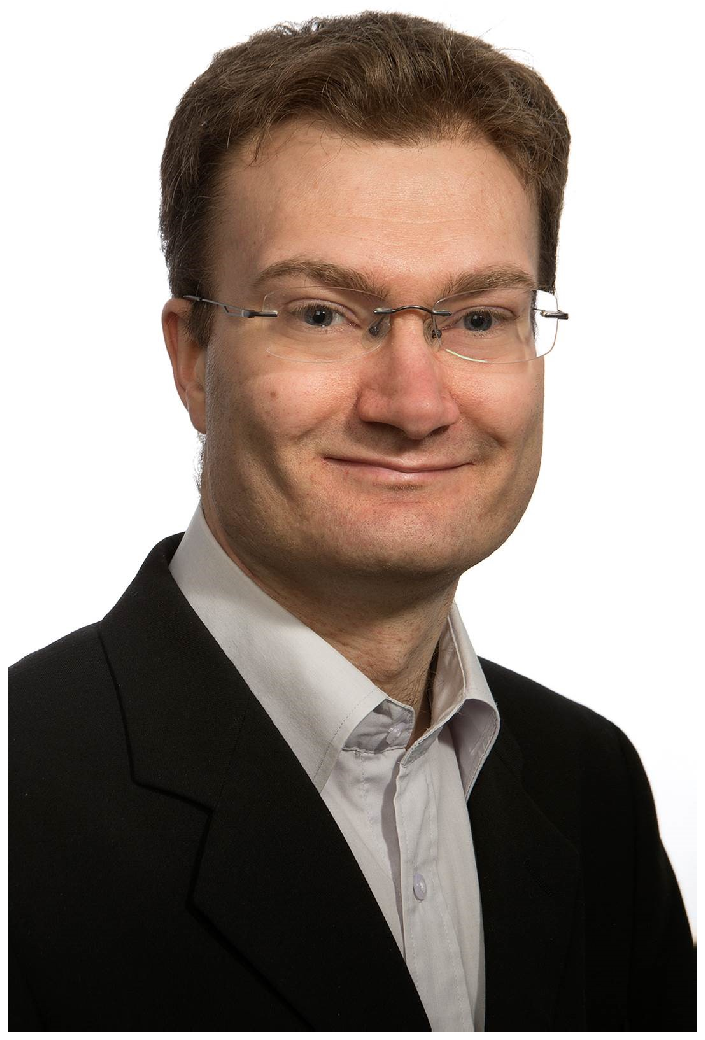}}]{Tor Inge Reigstad}
received the M.Sc degree from Department of Electric Power Engineering at the Norwegian University of Science and Technology (NTNU), Trondheim,
Norway, in 2007.
From 2007 to 2014, he was with Siemens AS, Trondheim, working with offshore grid power system analysis, magnetic component design and power converter development. From 2014 to 2017, he was with SINTEF Energy Research, Trondheim, where he was working with control of power grid and design and simulation of power electronic converters.
In 2018 he started his PhD studies within grid integration of variable speed hydro power.
His current research interests are mainly related to analysis and control of power electronic converters in power systems for variable speed hydro power.
\end{IEEEbiography}

\begin{IEEEbiography}[{\includegraphics[width=1in,height=1.25in,clip,keepaspectratio]{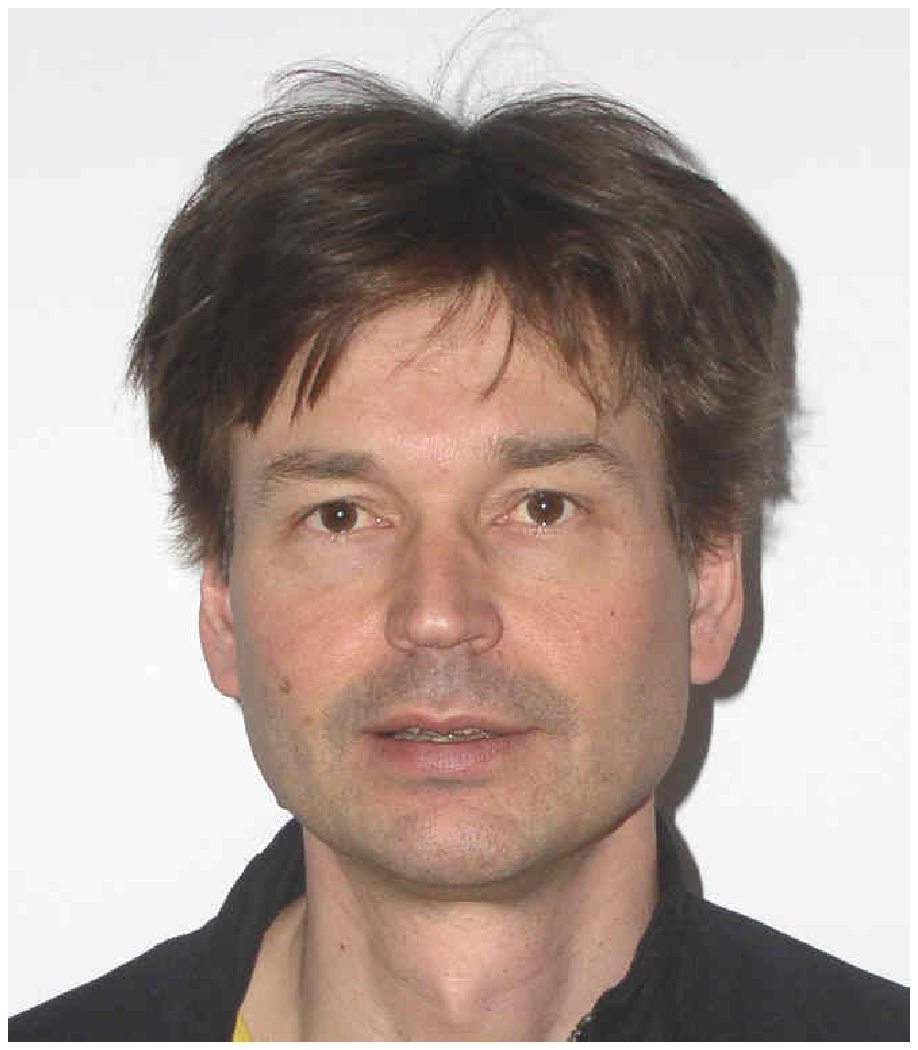}}]{Kjetil Uhlen}
is professor in Power Systems at the Norwegian University of Science and Technology (NTNU), Trondheim, and a Special Adviser at STATNETT (the Norwegian TSO). He has a Master degree (1986) and PhD degree (1994) in control engineering. His main areas of work include research and education within control and operation of power systems, grid integration of renewable energy and power system dynamics. 
\end{IEEEbiography}







\end{document}